\documentclass[twocolumn,aps,amsmath,nofootinbib,superscriptaddress]{revtex4}
\usepackage{psfig}

\begin{document}

\title{Evidence for the Onset of Deconfinement\\
from Longitudinal Momentum Distributions?\\ 
Observation of the Softest Point of the Equation of State}

\author{Marcus Bleicher}

\address{Institut f\"ur Theoretische Physik,
J.~W.~Goethe Universit\"at, 60438 Frankfurt am Main,
Germany}

\begin{abstract}
We analyze longitudinal pion spectra from $E_{\rm lab}= 2A$~GeV to $\sqrt
{s_{\rm NN}}=200$~GeV within Landau's hydrodynamical model.
From the measured data on the widths of the pion rapidity spectra, 
we extract the sound velocity $c_s^2$ in the early stage of the reactions.
It is found that the sound velocity has a local minimum (indicating a 
softest point in the equation of state, EoS) at $E_{\rm beam}=30A$~GeV.
This softening of the EoS is compatible with the assumption of the formation of a 
mixed phase at the onset of deconfinement.
\end{abstract}

\maketitle\nopagebreak

Over the last years, a wealth of detailed data in the $20A-160A$~GeV 
energy regime has become available. 
The systematic study of these data revealed surprising (non-monotonous) 
structures in various observables around $30A$~GeV beam energy.
Most notable irregular structures in that energy regime include, 
\begin{itemize}
\item 
the sharp maximum in the K$^+/\pi^+$ ratio \cite{Afanasiev:2002mx,Gazdzicki:2004ef},
\item
a step in the transverse momentum excitation function (as seen through 
$\langle m_\perp\rangle -m_0$ ) \cite{Gazdzicki:2004ef,na49_blume},
\item
an apparent change in the pion per participant ratio \cite{Gazdzicki:2004ef} and
\item
increased ratio fluctuations (due to missing data at low energies it is unknown if this 
is a local maximum or an ongoing increase of the fluctuations) \cite{Roland:2005pr}.
\end{itemize}

It has been speculated, that these observation hint towards the onset of deconfinement
already at $30A$~GeV beam energy. Indeed, increased strangeness production \cite{Koch:1986ud} 
and enhanced fluctuations have long been predicted as a sign of QGP 
formation \cite{Bleicher:2000ek,Shuryak:2000pd,Heiselberg:2000ti,Muller:2001wj,Gazdzicki:2003bb,Gorenstein:2003hk} within different 
frameworks and observables.
The suggestion of an enhanced strangeness to entropy ratio ($\sim K/\pi$) as indicator for the onset of QGP formation 
was especially advocated in \cite{SMES}. Also  the  high and approximately
constant $K^\pm$ inverse slopes of the $m_T$ spectra above $\sim 30A$~GeV - the 'step' - was also found to be consistent
with the assumption of a parton $\leftrightarrow$ hadron phase transition at low SPS 
energies \cite{Gorenstein:2003cu,Hama:2004re}.  
Surprisingly, transport simulations (supplemented by recent lattice QCD (lQCD) calculations) 
have also suggested that partonic degrees of freedom might already lead to
visible effects at $\sim 30A$~GeV \cite{Weber98,MT-prl,Bratkovskaya:2004kv}. 
Finally, the comparison of the thermodynamic parameters $T$ and $\mu_B$
extracted from the transport models in the central overlap region
\cite{Bravina} with the experimental systematics on chemical
freeze-out configurations \cite{Braun-Munzinger:1996mq,Braun-Munzinger:1998cg,Cleymans} 
in the $T-\mu_B$ plane do also suggest that a first glimpse on a deconfined state might be possible
around $10A-30A$~GeV.

In this letter, we explore whether similar irregularities are also present in the
excitation function of longitudinal observables, namely rapidity distributions.
Here we will employ Landau's hydrodynamical model \cite{Fermi:1950jd,Landau:gs,Belenkij:cd,Shuryak:1972zq,Carruthers:ws,Carruthers:dw,Carruthers:1981vs}. 
This model entered the focus again after the most remarkable observation that 
the rapidity distributions at all investigated energies can be well 
described by a single Gaussian at each energy. The energy dependence of the width
can also be reasonably described by the same model.
For recent applications of Landau's model to relativistic hadron-hadron and
nucleus-nucleus interactions the reader is referred to
\cite{Feinberg:1988et,Stachel:1989pa,Steinberg:2004vy,Murray:2004gh,Roland:2004} (and Refs. therein).

The main physics assumptions of Landau's
picture are as follows: The collision of two Lorentz-contracted nuclei 
leads to full thermalization in a volume of size
$V/\sqrt{s_{\rm NN}}$. This justifies the use of thermodynamics and
establishes the system size and energy dependence. Usually, a simple equation
of state  $p=c_s^2\epsilon$ with $c_s^2=1/3$ ($c_s$ denotes the speed of sound) 
is assumed. For simplicity, 
chemical potentials are not taken into account.
From these assumptions follows a  universal formula for the distribution of the produced entropy, determined mainly
by the initial Lorentz contraction and Gaussian rapidity spectrum
for newly produced particles. Under the condition that $c_s$ is independent of temperature, 
the rapidity density is given by \cite{Shuryak:1972zq,Carruthers:dw}:
\begin{equation}
\frac{dN}{dy}=\frac{Ks_{\rm NN}^{1/4}}{\sqrt{2\pi \sigma_y^2}}\,\exp\left(-\frac{y^2}{2\sigma_y^2}\right)
\label{eq1}
\end{equation}
with
\begin{equation}
\sigma_y^2=\frac{8}{3}\frac{c_s^2}{1-c_s^4}\,{\rm ln}({\sqrt {s_{\rm NN}}}/{2m_p})\quad,
\label{eq2}
\end{equation}
where $K$ is a normalisation factor and $m_p$ is the proton mass.
The model  relates the observed particle
multiplicity and distribution in a simple and direct way to the  parameters of 
the QCD matter under consideration.

\begin{figure}[t]
\vspace*{-.8cm}\psfig{file=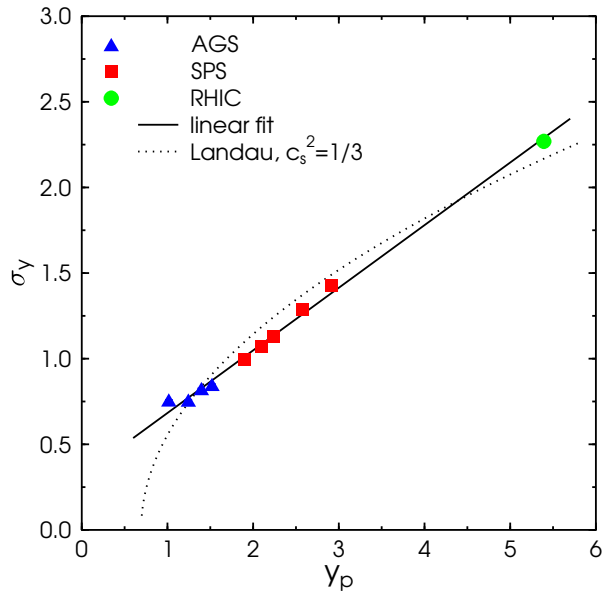,width=9.8cm} 
\vspace*{-.6cm}\caption{\label{rapwidth} The root mean square $\sigma_y$ of the rapidity
distributions of negatively charged pions in central Pb+Pb (Au+Au) reactions 
as a function of the beam rapidity $y_p$.
The dotted line indicates the Landau model prediction with  $c_s^2=1/3$, while
the full line shows a linear fit through the data points.
Data (full symbols) are taken from \cite{na49_blume,Roland:2004,klay,brahms}.
The statistical errors given by the experiments are smaller than the symbol sizes. 
Systematic errors are not available.}
\end{figure}

Let us now analyze the available experimental data on rapidity distributions of negatively 
charged pions in terms of the Landau model.
Fig. \ref{rapwidth} shows the measured root mean square $\sigma_y$ of the rapidity
distribution of negatively charged pions in central Pb+Pb (Au+Au) reactions 
as a function of the beam rapidity. The dotted line indicates 
the Landau model predictions with  the commonly used constant sound velocity $c_s^2=1/3$. 
The full line shows a linear fit through the data points, while the
data points \cite{na49_blume,Roland:2004,klay,brahms} are depicted by full symbols.
%%%%%%%%%%%%%%%%%%%%%%%%%%%%%%%%%%%%%%%%%%%%%%%%%%%%%%%%%%%%%%%%%%
%%%%%%%%%%%%%%%%%%%%%%%%%%%%%%%%%%%%%%%%%%%%%%%%%%%%%%%%%%%%%%%%%%5

At a first glance the energy dependence looks structureless.
The data seem to follow a linear dependence on the beam rapidity $y_p$ without
any irregularities.
However, the general trend of the rapidity widths is also well reproduced by 
Landau's model with an equation of state with a fixed speed of sound. 
Nevertheless, there seem to be systematic deviations.
At low AGS energies and at RHIC, the experimental points are generally
underpredicted by Eq.\ (\ref{eq2}), while in the SPS energy regime Landau's model overpredicts the
widths of the rapidity distributions.
Exactly these deviations from the simple Landau picture do allow to 
gain information on the equation of state 
of the matter produced in the early stage of the reaction.
By inverting Eq.\ (\ref{eq2}) we can express the speed of sound $c_s^2$ in the medium as a function of 
the measured width of the rapidity distribution:
\begin{equation}
c_s^2=-\frac{4}{3}\frac{{\rm ln}({\sqrt {s_{\rm NN}}}/{2 m_p})}{\sigma_y^2}
+\sqrt{\left[\frac{4}{3}\frac{{\rm ln}({\sqrt {s_{\rm NN}}}/{2 m_p})}{\sigma_y^2}\right]^2+1}\quad.
\label{eq3}
\end{equation}
%%%%%%%%%%%%%%%%%%%%%%%%%%%%%%%%%%%%%%%%%%%%%%%%%%%%%%%%%%%%%%%%%%
\begin{figure}[h!]
\vspace*{-.8cm}\psfig{file=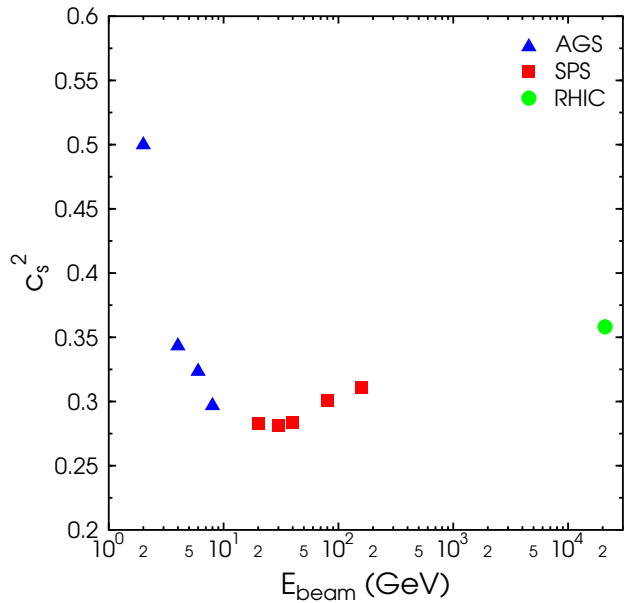,width=9.8cm} 
\vspace*{-.6cm}\caption{\label{cs2} Speed of sound as a function of beam energy for central 
Pb+Pb (Au+Au) reactions as extracted from the data using Eq.\ (\ref{eq3}).
The statistical errors (not shown) are smaller than 3\%.}
\end{figure}
%%%%%%%%%%%%%%%%%%%%%%%%%%%%%%%%%%%%%%%%%%%%%%%%%%%%%%%%%%%%%%%%%%5

Let us now investigate the energy dependence of the sound velocities extracted
from the data. Figure \ref{cs2} shows the speed of sound as a function of beam energy for central 
Pb+Pb (Au+Au) reactions as obtained from the data using Eq.\ (\ref{eq3}).
The sound velocities exhibit a clear  minimum (usually called the softest point) around a beam energy of
$30A$~GeV.
A localized softening of the equation of state is a long predicted  signal for the mixed phase 
at the transition energy from hadronic to partonic matter \cite{Hung:1994eq,Rischke:1995pe,Brachmann:1999mp}. 
Therefore, we conclude that the measured data on the rapidity widths of negatively charged pions
are indeed compatible with the assumption of the onset of deconfinement at the lower SPS energy range.
However, presently we can not rule out that also an increased resonance contribution may
be the cause of the softening \cite{soft}.

In conclusion, we have explored the excitation functions of the rapidity widths of negatively charged
pions in Pb+Pb (Au+Au) collisions.
\begin{itemize}
\item
The rapidity spectra of pions produced in  central nucleus-nucleus reactions at all investigated energies can be 
well described by single Gaussians.
\item
The energy dependence of the width of the pion rapidity distribution follows the
prediction of  Landau's hydrodynamical model if a variation of the sound
velocity  is taken into account.
\item
The speed of sound excitation function extracted from the data has a pronounced 
minimum (softest point) at $E_{\rm beam}=30A$~GeV.
\item
This softest point might be due to  the formation of a mixed phase indicating the onset of deconfinement at this energy.
\end{itemize}
Further explorations of this energy domain is needed and can be done at the future FAIR facility and 
by CERN-SPS and BNL-RHIC experiments.

\noindent
\section*{Acknowledgements}
The author thanks C. Blume and M. Gazdzicki for fruitful and stimulating discussions.
This work was supported by GSI, DFG and BMBF.
This work used computational resources provided by the
Center for Scientific Computing at Frankfurt (CSC).


\section*{References}
\begin{thebibliography}{10}

%\cite{Afanasiev:2002mx}
\bibitem{Afanasiev:2002mx}
  S.~V.~Afanasiev {\it et al.}  [The NA49 Collaboration],
  %``Energy dependence of pion and kaon production in central Pb + Pb
  %collisions,''
  Phys.\ Rev.\ C {\bf 66} (2002) 054902
  [arXiv:nucl-ex/0205002].
  %%CITATION = NUCL-EX 0205002;%%

%\cite{Gazdzicki:2004ef}
\bibitem{Gazdzicki:2004ef}
  M.~Gazdzicki {\it et al.}  [NA49 Collaboration],
  %``Report from NA49,''
  J.\ Phys.\ G {\bf 30} (2004) S701
  [arXiv:nucl-ex/0403023].
  %%CITATION = NUCL-EX 0403023;%%

\bibitem{na49_blume} C. Blume, J. Phys. G: Nucl. Part. Phys. 31, S57 (2005) 

%\cite{Roland:2005pr}
\bibitem{Roland:2005pr}
  C.~Roland  [NA49 Collaboration],
  %``Event-by-event fluctuations of particle ratios in central Pb + Pb
  %collisions at 20-A-GeV - 158-A-GeV,''
  J.\ Phys.\ G {\bf 31} (2005) S1075.
  %%CITATION = JPHGB,G31,S1075;%%

%\cite{Koch:1986ud}
\bibitem{Koch:1986ud}
  P.~Koch, B.~Muller and J.~Rafelski,
  %``Strangeness In Relativistic Heavy Ion Collisions,''
  Phys.\ Rept.\  {\bf 142} (1986) 167.
  %%CITATION = PRPLC,142,167;%%

%\cite{Bleicher:2000ek,Shuryak:2000pd,Heiselberg:2000ti,Muller:2001wj}
\bibitem{Bleicher:2000ek}
  M.~Bleicher, S.~Jeon and V.~Koch,
  %``Event-by-event fluctuations of the charged particle ratio from
  %non-equilibrium transport theory,''
  Phys.\ Rev.\ C {\bf 62} (2000) 061902
  [arXiv:hep-ph/0006201].
  %%CITATION = HEP-PH 0006201;%%

%\cite{Shuryak:2000pd}
\bibitem{Shuryak:2000pd}
  E.~V.~Shuryak and M.~A.~Stephanov,
  %``When can long range charge fluctuations serve as a QGP signal?,''
  Phys.\ Rev.\ C {\bf 63} (2001) 064903
  [arXiv:hep-ph/0010100].
  %%CITATION = HEP-PH 0010100;%%

%\cite{Heiselberg:2000ti}
\bibitem{Heiselberg:2000ti}
H.~Heiselberg and A.~D.~Jackson,
%``Anomalous multiplicity fluctuations from phase transitions in heavy ion
%collisions,''
Phys.\ Rev.\ C {\bf 63} (2001) 064904
[arXiv:nucl-th/0006021].
%%CITATION = NUCL-TH 0006021;%%

%\cite{Muller:2001wj}
\bibitem{Muller:2001wj}
  B.~Muller,
  %``Statistical fluctuations as probes of dense matter,''
  Nucl.\ Phys.\ A {\bf 702} (2002) 281
  [arXiv:nucl-th/0111008].
  %%CITATION = NUCL-TH 0111008;%%

  %\cite{Gazdzicki:2003bb,Gorenstein:2003hk}
  \bibitem{Gazdzicki:2003bb}
  M.~Gazdzicki, M.~I.~Gorenstein and S.~Mrowczynski,
  %``Fluctuations and deconfinement phase transition in nucleus nucleus
  %collisions,''
  Phys.\ Lett.\ B {\bf 585}, 115 (2004)
  [arXiv:hep-ph/0304052].
  %%CITATION = HEP-PH 0304052;%%

%  -kaon/pion ratio:

  %\cite{Gorenstein:2003hk}
  \bibitem{Gorenstein:2003hk}
  M.~I.~Gorenstein, M.~Gazdzicki and O.~S.~Zozulya,
  %``Fluctuations of strangeness and deconfinement phase transition in  nucleus
  %nucleus collisions,''
  Phys.\ Lett.\ B {\bf 585}, 237 (2004)
  [arXiv:hep-ph/0309142].
  %%CITATION = HEP-PH 0309142;%%


\bibitem{SMES}
    M. Gazdzicki and M. I. Gorenstein,
    Acta Phys. Polon. B {\bf 30}, 2705 (1999).
%\cite{Hama:2004re}

 %\cite{Gorenstein:2003cu}
  \bibitem{Gorenstein:2003cu}
  M.~I.~Gorenstein, M.~Gazdzicki and K.~A.~Bugaev,
  %``Transverse activity of kaons and the deconfinement phase transition in
  %nucleus nucleus collisions,''
  Phys.\ Lett.\ B {\bf 567}, 175 (2003)
  [arXiv:hep-ph/0303041].
  %%CITATION = HEP-PH 0303041;%%

\bibitem{Hama:2004re}
  Y.~Hama, F.~Grassi, O.~Socolowski, T.~Kodama, M.~Gazdzicki and M.~Gorenstein,
  %``Energy dependence of the inverse slope parameter in heavy ion collisions,''
  Acta Phys.\ Polon.\ B {\bf 35} (2004) 179.
  %%CITATION = APPOA,B35,179;%%

\bibitem{Weber98}
    H. Weber, C. Ernst, M. Bleicher {\it et al.},
    Phys. Lett. B {\bf 442}, 443 (1998).

\bibitem{MT-prl}
    E. L. Bratkovskaya {\it et al.},
    Phys. Rev. Lett. {\bf 92}, 032302 (2004)

\bibitem{Bratkovskaya:2004kv}
E.~L.~Bratkovskaya {\it et al.},
%``Strangeness dynamics and transverse pressure in relativistic nucleus nucleus
%collisions,''
Phys.\ Rev.\ C {\bf 69}, 054907 (2004)
%[arXiv:nucl-th/0402026].
%%CITATION = NUCL-TH 0402026;%%

\bibitem{Bravina}
    L. V. Bravina {\it et al.}, Phys. Rev. C {\bf 60}, 024904 (1999),
     Nucl. Phys. A {\bf 698}, 383 (2002).

%\cite{Braun-Munzinger:1996mq,Braun-Munzinger:1998cg}
\bibitem{Braun-Munzinger:1996mq}
  P.~Braun-Munzinger and J.~Stachel,
  %``Probing the phase boundary between hadronic matter and the
  %quark-gluon-plasma in relativistic heavy ion collisions,''
  Nucl.\ Phys.\ A {\bf 606}, 320 (1996).
  %%CITATION = NUCL-TH 9606017;%%

%\cite{Braun-Munzinger:1998cg}
\bibitem{Braun-Munzinger:1998cg}
  P.~Braun-Munzinger and J.~Stachel,
  %``Dynamics of ultra-relativistic nuclear collisions with heavy beams: An
  %experimental overview,''
  Nucl.\ Phys.\ A {\bf 638}, 3 (1998).
  %%CITATION = NUCL-EX 9803015;%%

\bibitem{Cleymans}
    J. Cleymans and K. Redlich, Phys. Rev. C {\bf 60}, 054908 (1999).


\bibitem{Fermi:1950jd} E.~Fermi, Prog.\ Theor.\ Phys.\  {\bf 5}, 570 (1950).

\bibitem{Landau:gs} L.~D.~Landau, Izv.\ Akad.\ Nauk Ser.\ Fiz.\  {\bf 17}, 51 (1953).

\bibitem{Belenkij:cd} S.~Z.~Belenkij and L.~D.~Landau, Usp.\ Fiz.\ Nauk {\bf 56}, 309 (1955).

%\cite{Shuryak:1972zq}
\bibitem{Shuryak:1972zq}
  E.~V.~Shuryak,
  %``Multiparticle Production In High Energy Particle Collisions,''
  Yad.\ Fiz.\  {\bf 16}, 395 (1972).
  %%CITATION = YAFIA,16,395;%%


\bibitem{Carruthers:dw} P.~Carruthers, Annals N.Y.Acad.Sci. 229, 91 (1974).

\bibitem{Carruthers:ws} P.~Carruthers and M.~Doung-van, Phys.\ Rev.\ D {\bf 8}, 859 (1973).

\bibitem{Carruthers:1981vs} P.~Carruthers, LA-UR-81-2221 

%\cite{Feinberg:1988et}
\bibitem{Feinberg:1988et}
  E.~L.~Feinberg,
  %``Initial State Formation In The Hydrodynamical Theory Of Nucleon Nucleon
  %Collisions,''
  Z.\ Phys.\ C {\bf 38} (1988) 229.
  %%CITATION = ZEPYA,C38,229;%%

%\cite{Stachel:1989pa}
\bibitem{Stachel:1989pa}
  J.~Stachel and P.~Braun-Munzinger,
  %``Stopping In High-Energy Nucleus Nucleus Collisions: Analysis In The Landau
  %Hydrodynamic Model,''
  Phys.\ Lett.\ B {\bf 216},  1 (1989).
  %%CITATION = PHLTA,B216,1;%%

%\cite{Steinberg:2004vy}
\bibitem{Steinberg:2004vy}
P.~Steinberg,
%``Landau hydrodynamics and RHIC phenomena,''
arXiv:nucl-ex/0405022.
%%CITATION = NUCL-EX 0405022;%%

\bibitem{Murray:2004gh} M.~Murray, arXiv:nucl-ex/0404007.

\bibitem{Roland:2004} G. Roland, Talk presented at Quark Matter 2004, see proceedings.

\bibitem{klay}
J.Klay {\it et al.} [E895 Collaboration], Phys.\ Rev.\ C {\bf 68}, 054905 (2003)

\bibitem{brahms}
I.G. Bearden {\it et al.} [Brahms Collaboration], Phys.\ Rev.\ Lett. {\bf 94}, 162301 (2005)

%\cite{Hung:1994eq}
\bibitem{Hung:1994eq}
  C.~M.~Hung and E.~V.~Shuryak,
  %``Hydrodynamics near the QCD phase transition: Looking for the longest lived
  %fireball,''
  Phys.\ Rev.\ Lett.\  {\bf 75}, 4003 (1995)
  [arXiv:hep-ph/9412360].
  %%CITATION = HEP-PH 9412360;%%

%\cite{Rischke:1995pe}
\bibitem{Rischke:1995pe}
  D.~H.~Rischke, Y.~Pursun, J.~A.~Maruhn, H.~Stoecker and W.~Greiner,
  %``The phase transition to the quark-gluon plasma and its effects on
  %hydrodynamic flow,''
  Heavy Ion Phys.\  {\bf 1}, 309 (1995) 
  [arXiv:nucl-th/9505014].
  %%CITATION = NUCL-TH 9505014;%%

%\cite{Brachmann:1999mp}
\bibitem{Brachmann:1999mp}
  J.~Brachmann, A.~Dumitru, H.~Stoecker and W.~Greiner,
  %``The directed flow maximum near c(s) = 0,''
  Eur.\ Phys.\ J.\ A {\bf 8}, 549 (2000) 
  [arXiv:nucl-th/9912014].
  %%CITATION = NUCL-TH 9912014;%%

\bibitem{soft}
 R. Hagedorn, Nuov. Cim. Suppl. 3, 147 (1965); J. Rafelski and R. Hagedorn, Bielefeld Symp., ed. H. Satz, pp. 253 (1980) 
\end{thebibliography}
\end{document}